\documentclass[12pt]{article}
\usepackage{amsmath,amssymb,latexsym,axodraw,a4wide}
\usepackage[dvips]{epsfig}
\usepackage[small]{caption}
\advance\topmargin -1.5cm
\advance\textheight 2cm

\newcommand{\dr}{differential renormalization}

\let\dots\textellipsis
\def\limfunc#1{\mathop{\rm #1}}

\def\Li2{\limfunc{Li_2}}

\def\ie{i.e.\ }
\def\eg{e.g.\ }
\def\MW{M_{\rm W}}
\def\MZ{M_{\rm Z}}
\def\MH{M_{\rm H}}
\def\me{m_{\rm e}}
\def\pslash#1{\rlap{\kern .1ex/}#1}
\def\unity{{\rm 1\mskip-4.25mu l}}

\newcommand{\FA}{{\sl FeynArts}}
\newcommand{\mma}{{\sl Mathematica}}
\newcommand{\FC}{{\sl FormCalc}}
\newcommand{\FO}{{\sl FORM}}
\newcommand{\LT}{{\sl LoopTools}}
\newcommand{\WWWW}{$\mathrm{W^+W^-\to W^+W^-}$}
\newcommand{\ZZZZ}{$\mathrm{ZZ\to ZZ}$}

\newcommand{\cpc}[3]{{\sl Comp. Phys. Commun.} {\bf #1} (19#2) #3}
\newcommand{\fp}[3]{{\sl Fortschr. Phys.} {\bf #1} (19#2) #3}

\newcommand{\np}[3]{{\sl Nucl. Phys.} {\bf #1} (19#2)~#3}
\newcommand{\nim}[3]{{\sl Nucl. Instr. Meth.} {\bf #1} (19#2)~#3}
\newcommand{\pl}[3]{{\sl Phys. Lett.} {\bf #1} (19#2) #3}
\newcommand{\pr}[3]{{\sl Phys. Rev.} {\bf #1} (19#2) #3}

\newcommand{\zp}[3]{{\sl Z. Phys.} {\bf #1} (19#2) #3}
\newcommand{\vj}[4]{{\sl #1} {\bf #2} (19#3) #4}

\begin{document}

\thispagestyle{empty}
\def\thefootnote{\fnsymbol{footnote}}
\setcounter{footnote}{1}
\null
\strut\hfill UG-FT-87/98\\
\strut\hfill KA-TP-7-1998\\
\strut\hfill hep-ph/9807565
\vskip 0cm
\vfill
\begin{center}
{\Large\bf Automatized One-Loop Calculations in}\\[1ex]
{\Large\bf four and \textit{D} dimensions}
\vskip 2.5em
{\large\sc T.~Hahn}\\[1ex]
{\it Institut f\"ur Theoretische Physik, Universit\"at Karlsruhe\\
D-76128 Karlsruhe, Germany}\\[2ex]
{\large\sc M.~P\'erez-Victoria}\\[1ex]
{\it Dpto.\ de F\'{\i}sica Te\'orica y del Cosmos, Universidad de Granada\\
E-18071 Granada, Spain}
\par \vskip 1em
\end{center} \par
\vskip 1cm 
\vfill
{\bf Abstract:}\\
Two program packages are presented for evaluating one-loop amplitudes. 
They can work either in dimensional regularization or in constrained
differential renormalization. The latter method is found at the one-loop
level to be equivalent to regularization by dimensional reduction.
\par
\vskip 1cm 
\noindent
July 1998\par
\null
\setcounter{page}{0}
\clearpage
\def\thefootnote{\arabic{footnote}}
\setcounter{footnote}{0}


\section{Introduction}

Little needs to be said nowadays about the importance of one-loop
calculations. During the last years methods which (partially) automatize
these calculations have become available \cite{MeBD91, grace, xloops}.
The automatization is though nowhere as fully developed as for tree-level
calculations \cite{comphep, madgraph}. The reason is that loop
calculations are in general more involved, and moreover a complete
automatization limits the possibilities to use a program beyond its
designed scope. 

In virtually all implementations of one-loop calculations on the computer,
dimensional regularization \cite{tHV72,otherdimreg} has been employed for
calculating renormalized amplitudes, and not without reason as it is a
very elegant formalism that preserves most symmetries. However,
dimensional regularization presents problems in chiral theories, related
to the extension of $\gamma_5$ to arbitrary dimensions
\cite{AD73,BGL72,BM77,Bo80}. In the Standard Model, a naive
scheme---anticommuting $\gamma_5$ without further modifications---is
usually used, but the situation is not completely clear \cite{CFH79}. More
problematic are supersymmetric theories, where a variant, regularization
by dimensional reduction \cite{Si79}, is commonly used in spite of
possible inconsistencies at higher loops \cite{Si80}.  On the other hand,
a new method in 4 dimensions, constrained differential renormalization
(CDR), has recently been developed \cite{CDR,techniques}. It is a version
of \dr\ \cite{FJL} that preserves gauge invariance at least to one loop
\cite{CDR,techniques,nonabelian}. There are some hints that it also
preserves supersymmetry \cite{SUSY}. At any rate, it is clearly convenient
to have at hand alternative methods implemented for automatized one-loop
calculations.

We have implemented the three mentioned methods in two program packages,
called \FC\ and \LT, for the evaluation of one-loop amplitudes. It turns
out that CDR and dimensional reduction are equivalent at the one-loop
level (at least in the cases where the latter is well-defined). This is a
new result and will be discussed in the next section. Actually, once the
CDR coordinate-space expressions and procedure are translated into
momentum space, both methods can be set up in an identical way. Therefore,
our programs have in fact only two different options: to perform
calculations in dimensional regularization or in dimensional
reduction~/~CDR. We shall refer to these two possibilities as calculating
in $D$ or 4 dimensions, respectively. 

Our programs do not fully automatize one-loop calculations to the extent
that they directly produce cross-sections from the process specification.
Instead, \FC\ produces output which can easily be evaluated further, \eg
numerically in a Fortran program. One big advantage of \FC\ is that it
leaves the user with a \mma\ expression which is considerably easier to
modify than Fortran code. \LT, on the other hand, supplies all the
functions needed for the numerical evaluation of the \FC\ output. We
supply also a demonstration program to show how this numerical evaluation
can be done. 

The paper is organized as follows: In Section \ref{sect:methods} we
outline the regularization methods we have used in the programs, and
discuss the equivalence between dimensional reduction and CDR. In Section
\ref{sect:computer} the implementation of the program packages is
explained, where Section \ref{sect:formcalc} describes the analytical and
Section \ref{sect:looptools} the numerical part. In Section
\ref{sect:example} we illustrate for the example of elastic Z--Z
scattering the individual steps of a one-loop calculation using our
packages. Section \ref{sect:availability} lists the computer requirements
needed to compile and run the packages, and their availability. Finally,
an appendix collects the functions implemented for numerical evaluation.

\section{Calculations in 4 and \textit{D} dimensions}
\label{sect:methods}

{\bf\textit{Dimensional regularization}} has proved to be the most
convenient method for computing quantum corrections in gauge theories. The
definition of the regularized expressions has two parts: analytic
continuation of momenta (and other four-vectors) in the number of
dimensions, $D$, and an extension to $D$ dimensions of the Lorentz
covariants ($\gamma_\mu$, $g_{\mu\nu}$, etc.). The second part is achieved
by treating the covariants as formal objects obeying certain algebraic
identities \cite{tHV72}. The dimensionally regularized Feynman graphs 
are meromorphic functions of the complex parameter $D$, and the poles at 
$D = 4$ can consistently be subtracted to all orders in perturbation theory.
Such a minimal subtraction scheme (MS) defines a renormalization
fulfilling the usual requirements of a quantum field theory: causality,
unitarity, field equations, Ward identities, etc.\ \cite{tHV72,BM77}. 
Problems only appear for identities that depend on the 4-dimensional 
nature of the objects involved. This is the case for the Fierz identities or for
relations using the Levi-Civita tensor. In particular, the extension of
$\gamma_5$ to $D$ dimensions is problematic. The consistent prescriptions
of Refs.\ \cite{tHV72,AD73,BM77} lead to spurious anomalies that have to
be corrected with finite counterterms. We follow instead Ref.\
\cite{CFH79} and work with an anticommuting $\gamma_5$. We also maintain
the usual relations for traces, such as
$\limfunc{Tr}(\gamma_5\gamma_\mu\gamma_\nu\gamma_\rho\gamma_\sigma) =
\varepsilon_{\mu\nu\rho\sigma}\mathrm{Tr}\unity$. Even though
they can be incompatible with an anticommuting $\gamma_5$ in $D$
dimensions, the resulting ambiguities are expected to cancel in
non-anomalous theories \cite{CFH79,Sh89}. For one-loop calculations one
can proceed in the following manner (see Ref.\ \cite{De93} for details):
\begin{enumerate}
\item	Calculate traces and simplify the Dirac algebra in $D$
	dimensions.
\item	Write everything in terms of scalar and tensor integrals in $D$
	dimensions.
\item	Decompose tensor integrals into Lorentz-covariant tensors
	constructed from the external momenta and the metric tensor. One
	is left with scalar integrals as coefficients of these
	tensors.
\item	Calculate the scalar integrals and tensor coefficients in
	$D$ dimensions, expand the whole diagram in 
	$\varepsilon = D - 4$, subtract the $\frac 1\varepsilon$ poles 
	and take the limit $\varepsilon \rightarrow 0$. Computer
	algebraically it is convenient to do this in two steps:
	\begin{enumerate}
	\item	Add local terms for products of $D$ times a divergent 
		integral: $D I \rightarrow 4 I + c$, with $I$ the 
		integral and $c$ the coefficient of its 
		$\frac 1\varepsilon$ pole \label{localterms}. 
		Then, make $D=4$ outside the integrals.
	\item	Calculate the scalar integrals and tensor 
		coefficients in $D$ dimensions, expand them up to 
		order $\varepsilon^0$, and subtract their poles.
	\end{enumerate}
\end{enumerate}
A modified version of dimensional regularization, designed to preserve
supersymmetry and gauge invariance, was proposed by Siegel \cite{Si79}. 
It is based on {\bf\textit{dimensional reduction}} from 4 to $D$ 
dimensions:
while the integration momenta are $D$-dimensional, as in usual dimensional
regularization, all other tensors and spinors are kept 4-dimensional.
Therefore one works with two kinds of objects: $D$-dimensional and
4-dimensional ones. For the validity of the field equations and gauge
invariance, one must impose the identity 
\begin{equation}
\label{dimredeq}
g_{\mu\nu} \hat{g}^{\nu\rho} = \hat{g}_\mu^{~\rho}\,,
\end{equation}
where $g_{\mu\nu}$ is the 4-dimensional metric ($g^\mu{}_\mu=4$) and
$\hat g_{\mu\nu}$ the $D$-dimensional one ($\hat g^\mu{}_\mu=D$). The
Dirac algebra, including $\gamma_5$, is performed in 4 dimensions.
Unfortunately, regularization by dimensional reduction is known to be
inconsistent \cite{Si80}. Nevertheless the inconsistencies arise at higher
orders and the method has successfully been applied to many calculations
in supersymmetric theories.\footnote{%
Dimensional reduction can also be used in non-supersymmetric theories, but
more care is required in order not to break unitarity \cite{CJN80}.}
At one loop, the procedure described for dimensional regularization can be
followed. The only differences are that the Dirac algebra is 4-dimensional
and that a 4-dimensional metric tensor and Eq.\ (\ref{dimredeq}) have to
be introduced. Notice that $\hat g_{\mu\nu}$ only appears in the
decomposition of the tensor integrals. These (and the tensor coefficients)
are identical in dimensional regularization and dimensional reduction. 

Finally, {\bf\textit{\dr}} is a method of regularization and
renormalization in coordinate space that cures UV divergences by
substituting badly-behaved expressions by derivatives of well-behaved
ones. The method has proved to be quite simple and convenient in a number
of applications. A symmetric procedure of \dr\ preserving gauge invariance
at the one-loop level has recently been proposed.  This so-called
constrained \dr\ proceeds in two steps:
\begin{enumerate}
\item	The expressions read from Feynman diagrams are written in terms
	of a complete set of (singular) {\em basic functions}. The basic
	functions are products of propagators with derivatives acting on
	one of them. Functions are treated differently depending on
	whether their indices are self-contracted or not. All algebra is
	performed strictly in 4 dimensions.

\item	These basic functions are substituted by their {\em renormalized}
	expressions. The renormalization of the singular basic functions
	has previously been fixed once and for all, such that a set of
	simple rules is respected (these rules are just natural
	extensions of mathematical identities among tempered
	distributions) \cite{techniques}. In particular, the rules imply
	that renormalization does not commute with index contraction.
\end{enumerate}
Although \dr\ naturally works in coordinate space, it is possible to
perform the reduction of Step 1 in momentum space and use the Fourier
transforms of the renormalized basic functions in Step 2, which correspond
to the tensor integrals in the dimensional methods. Expressions and
manipulations in coordinate space can easily be translated into momentum
space (see Appendix~B of Ref.\ \cite{techniques}).

As mentioned in the introduction, it turns out that CDR and dimensional
reduction are equivalent methods for one-loop calculations. The
equivalence can easily be understood once CDR has been translated into
momentum space. First, the minimally subtracted $D$-dimensional tensor
integrals are identical in the limit $D\to 4$ to the Fourier transforms of
the corresponding renormalized basic functions of CDR (which we shall also
call tensor integrals in the following), up to a redefinition of the
renormalization scale.\footnote{%
We have explicitly checked that the CDR tensor integrals in Appendix~B
of Ref.\ \cite{techniques} coincide with the tensor integrals in
dimensional reduction in the $\overline{\text{MS}}$ scheme if the
CDR renormalization scale $\bar M$ is redefined as $\log \bar M^2 =
\log \mu^2 + 2$ where $\mu$ is the renormalization scale in the
dimensional method.}
The reason is that one-loop integrals in $D$ dimensions satisfy the
relations imposed by the CDR rules. Hence, a possible discrepancy can only
arise in the initial conditions (namely, the renormalized value of the
scalar one- and two-point functions), but this can be taken care of by the
freedom to choose the renormalization scale. Second, all the algebra
outside tensor integrals is considered 4-dimensional in both schemes.
Third, the distinction of tensor integrals with contracted and
uncontracted indices in CDR is imitated in dimensional reduction by Eq.\
(\ref{dimredeq}) and the addition of local terms in
Step~\ref{localterms}. To illustrate this, consider the renormalized
tensor integral $C_{\mu\nu}$ (see Ref.~\cite{De93} for its definition),
which is the same in both methods (up to the freedom to choose the
renormalization scheme). In CDR the tensor integral $C^\mu{}_\mu$ is
considered independently; its renormalized value differs from $g^{\mu\nu}$
times the renormalized $C_{\mu\nu}$ by a local term \cite{CDR}. In
dimensional reduction we have
\begin{equation}
\begin{aligned}
g^{\mu\nu} C_{\mu\nu}
&= g^{\mu\nu} \left(\hat g_{\mu\nu} C_{00} +
\sum_{i,j=1}^2 p_{i\mu} p_{j\nu} C_{ij} \right) 
= \hat g^\mu{}_\mu C_{00} + \sum_{i,j=1}^2 (p_i p_j) C_{ij}
\displaybreak[0] \\
&= D C_{00} + \sum_{i,j=1}^2 (p_i p_j) C_{ij}
= 4 C_{00} - \frac 12 + \sum_{i,j=1}^2 (p_i p_j) C_{ij} \, .
\end{aligned}
\end{equation}
$C_{00}$ and $C_{ij}$ coincide in both methods.  The extra local term,
$-1/2$, is precisely what is needed to obtain $C^\mu{}_\mu$ from
$g^{\mu\nu}C_{\mu\nu}$ in CDR \cite{CDR,techniques}. That the same occurs
in the general case follows from the fact that in dimensional reduction
one can contract the 4-dimensional metric with $D$-dimensional integration
momenta before performing the integrals, and the resulting contracted
tensor integrals also satisfy the CDR relations. Notice that conventional
dimensional regularization introduces extra $D$'s coming from 
$\hat g^\mu{}_\mu$'s outside the tensor integrals. Also, the Feynman
rules in dimensional regularization contain $D$ in some theories.
Hence it can render different results, and not only in intermediate steps.
A simple example where the results in $D$ and 4 dimensions differ is the
electron self-energy in QED:
\begin{align}
\label{eq:seDdim}
\lower 3.8ex\hbox{\epsfig{file=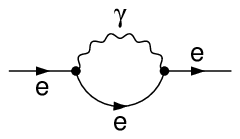, height=9ex}}
&\overset{\text{dim.reg.}}{=} -\frac{e^2}{16\pi^2}\left[
4 \me B_0(k^2, \me^2, 0) + 2 \pslash{k} B_1(k^2, \me^2, 0)
+ \pslash{k} - 2\me
\right]\,, \\
\label{eq:se4dim}
&\overset{\text{CDR,}}{\overset{\text{dim.red.}}{=}}
-\frac{e^2}{16\pi^2}\left[
4 \me B_0(k^2, \me^2, 0) + 2 \pslash{k} B_1(k^2, \me^2, 0)
\right]\,.
\end{align}
The functions $B_0$ and $B_1$ are defined in App.\ \ref{sect:reflt}.

Since dimensional reduction and CDR are equivalent at the one-loop level,
one only needs a single implementation for the two methods. We have
decided to use the CDR idea and add local terms when there are
self-contracted indices after all the simplifications have been performed.
This allows to work completely in 4 dimensions. Of course, the
dimensionally regularized scalar functions and tensor coefficients are
also used in this case.

\section{Implementation of the program packages} 
\label{sect:computer}

The evaluation of one-loop diagrams using our packages proceeds in two
steps:

In \FC\ the symbolic expressions for the diagrams as obtained from \FA\
\cite{KuBD91} are simplified algebraically such that the output can be
used almost directly in a numerical program, \eg in Fortran or C++.

\LT\ is the second of the packages and supplies the actual numerical
implementations of the one-loop functions needed for programs made from
the \FC\ output. It is based on the reliable package {\sl FF} \cite{vOV90}
and provides in addition to the scalar integrals of {\sl FF} also the
tensor coefficients in the conventions of Ref.\ \cite{De93}. \LT\ offers
three interfaces: Fortran, C++, and \mma, so most programming tastes
should be served.

\subsection{\FC}
\label{sect:formcalc}

\FC\ is a \mma-based program to calculate one-loop Feynman diagrams
in either CDR or dimensional regularization. The program reads
input from \FA\ and returns output in a way well suited for further
numerical (or analytical) evaluation. \FC\ was designed to handle large
numbers of diagrams, for example running the complete \WWWW\ amplitudes
(about 1000 diagrams \cite{DeH98}) takes approximately 10 minutes on a
Pentium PC.

The structure of \FC\ is rather simple: it prepares the symbolic
expressions of the diagrams in an input file for \FO, runs \FO, and
retrieves the results. This interaction is transparent to the user. \FC\
combines the speed of \FO\ with the powerful instruction set of \mma\ and
the latter greatly facilitates further processing of the results. The
following diagram shows schematically how \FC\ interacts with \FO\/: 
\bigskip
\begin{center}
\setlength{\unitlength}{1pt}
\begin{picture}(310,100)
\SetWidth{1}
\EBox(5,20)(105,100)
\EBox(205,30)(305,90)
\SetWidth{.5}
\Line(5,75)(105,75)
\Line(5,55)(105,55)
\ArrowLine(105,72)(205,72)
\EBox(125,50)(185,66)
\ArrowLine(205,58)(185,58)
\ArrowLine(125,58)(105,58)
\Text(155,58)[]{\tt ReadForm}
\Text(55,88)[]{\FC}
\Text(55,64)[]{\tt OneLoop[{\it amp}]}
\Text(55,45)[]{\mma}
\Text(55,30)[]{\scriptsize (user friendly, slow)}
\Text(255,75)[]{\FO}
\Text(255,55)[]{\scriptsize (very fast, but}
\Text(255,43)[]{\scriptsize not so user friendly)}
\SetWidth{.3}
\Line(0,15)(0,10)
\Line(0,10)(108,10)
\Line(108,10)(108,15)
\Line(110,15)(110,10)
\Line(110,10)(310,10)
\Line(310,10)(310,15)
\Text(55,5)[t]{\it user interface}
\Text(210,5)[t]{\it internal \FC\ calls}
\end{picture}
\end{center}

The main function in \FC\ is {\tt OneLoop} (the name is not strictly
correct since it works also with tree graphs). It is used like
this:
\begin{verbatim}
   << FormCalc.m;
   $Dimension = 4;
   amps = << myamps.m;
   alldiags = OneLoop[amps]
\end{verbatim}
The file {\tt myamps.m} is assumed here to contain amplitudes generated
by \FA. The dimension---{\tt D} for dimensional regularization or {\tt 4}
for dimensional reduction / CDR---is set with {\tt \$Dimension}. 

Alternatively, if one wants to evaluate only a subset of diagrams,
\begin{verbatim}
   somediags = OneLoop[ Pick[amps, {3, 5, {21, 29}}] ]
\end{verbatim}
will calculate diagrams 3, 5, and 21 through 29, or for a single diagram
it can be just
\begin{verbatim}
   onediag = OneLoop[ amps[[9]] ]
\end{verbatim}
Note that {\tt OneLoop} needs no declarations of the kinematics of the
underlying process; it uses the information \FA\ hands down. When
amplitudes are loaded, \FC\ automatically puts the external particles on
shell and for some common processes\footnote{%
Currently processes with 4 or 5 external legs, \ie $2\to 2$, $2\to 3$,
$1\to 3$, $1\to 4$.}
introduces kinematical invariants (\eg Mandelstam variables for a $2\to 2$
process). The former can be disabled by setting {\tt \$OnShell = False}
before loading the amplitudes.

For example, the QED electron self-energy in Eq.\ (\ref{eq:seDdim}) was
calculated with the following short program:
\begin{verbatim}
   << FormCalc.m;
   $OnShell = False;
   $Dimension = D;
   OneLoop[<< electronSE.amp] //. Abbreviations[]
\end{verbatim}
with the result
\begin{tt}
\begin{multline*}
\text{-}\,
\frac{\text{EL$^{\text{2}}$ ME B0[Pair[k[1],\,k[1]], ME2, 0]}}
{\text{4 Pi$^{\text{2}}$}}
\:\text{-} \\[1.5ex]
\frac{\text{EL$^{\text{2}}$ B1[Pair[k[1],\,k[1]], ME2, 0] ga[k[1]]}}
{\text{8 Pi$^{\text{2}}$}}
\:\text{-}\:
\frac{\text{EL$^{\text{2}}$ (-2 ME + ga[k[1]])}}
{\text{16 Pi$^{\text{2}}$}}
\end{multline*}
\end{tt}
Similarly, Eq.\ (\ref{eq:se4dim}) can be reproduced by putting
{\tt \$Dimension = 4}. This results in
\begin{tt}
\begin{multline*}
\text{-}\,
\frac{\text{EL$^{\text{2}}$ ME B0[Pair[k[1],\,k[1]], ME2, 0]}}
{\text{4 Pi$^{\text{2}}$}}
\:\text{-} \\[1.5ex]
\frac{\text{EL$^{\text{2}}$ B1[Pair[k[1],\,k[1]], ME2, 0] ga[k[1]]}}
{\text{8 Pi$^{\text{2}}$}}
\hphantom{\:\text{-}\:
\frac{\text{EL$^{\text{2}}$ (-2 ME + ga[k[1]])}}
{\text{16 Pi$^{\text{2}}$}}}
\end{multline*}
\end{tt}

Even more comprehensive than {\tt OneLoop}, the function {\tt ProcessFile}
can process entire files. It splits the results up into a bosonic and a
fermionic part; this is commonly needed for the numerical evaluation \eg
if one wants to sum over fermion generations. {\tt ProcessFile} is invoked
\eg as
\begin{verbatim}
   ProcessFile["vertex.amp", "results/vertex"]
\end{verbatim}
It reads the \FA\ amplitudes from {\tt vertex.amp} and produces
{\tt results/vertex.m} which contains the bosonic part and
{\tt results/vertexF.m} which contains the fermionic part.

{\tt OneLoop} and {\tt ProcessFile} return expressions where spinor
chains, dot products of and Levi-Civita tensors contracted with
polarization vectors have been collected and abbreviated. A term in such
an expression may look like
\begin{verbatim}
   C0i[cc1, MW2, S, MW2, MZ2, MW2, MW2] *
     (P12*S*(-8*a2*MW2 + 4*a2*MW2*S2 - 28*a2*CW^2*MW2*S2 +
             16*a2*CW^2*S*S2 + 4*a2*C2*MW2*SW^2) +
      O47*S*(-32*a2*CW^2*MW2*S2 + 8*a2*CW^2*S2*T + 8*a2*CW^2*S2*U) -
      P13*S*(-64*a2*CW^2*MW2*S2 + 16*a2*CW^2*S2*T + 16*a2*CW^2*S2*U) +
      O3*(-2*a2*MW2*T + a2*MW2*S2*T - 7*a2*CW^2*MW2*S2*T +
          4*a2*CW^2*S*S2*T + a2*C2*MW2*SW^2*T + 2*a2*MW2*U -
          a2*MW2*S2*U + 7*a2*CW^2*MW2*S2*U - 4*a2*CW^2*S*S2*U -
          a2*C2*MW2*SW^2*U))
\end{verbatim}
The first line stands for the tensor coefficient function
$C_1(\MW^2, s, \MW^2, \MZ^2, \MW^2, \MW^2)$ (see Appendix) which is
multiplied with a linear combination of abbreviations like {\tt O47} or
{\tt P12} with certain coefficients.

These coefficients contain the Mandelstam variables {\tt S}, {\tt T},
and {\tt U} and some short-hands for parameters of the Standard Model,
\eg ${\tt a2} = \alpha^2$ or ${\tt C4} = \cos^{-4}\theta_{\rm W}$.
Replacing such commonly appearing factors by symbols makes expressions
faster to handle for \mma. The coefficients are collected in a way
optimized for numerical evaluation: for example, \mma\ will automatically
add together the terms in {\tt -2*a2*MW2*T + a2*MW2*S2*T -
7*a2*CW\symbol{94}2*MW2*S2*T} once the numerical values for the various
parameters are substituted (\ie the result is of the form {\tt (a
number)*T}), so it is not worth simplifying this expression
further. If desired, \FC\ can nevertheless perform even these last
simplifications to obtain the most compact analytical form.\footnote{%
\FC\ wraps two functions, {\tt o1} and {\tt o2}, around coefficients at
different levels in the amplitude. The default setting {\tt o1 = o2 =
Identity} can be changed \eg to {\tt o1 = o2 = Simplify}.}

The abbreviations like {\tt O47} or {\tt P12} are introduced automatically
and can significantly reduce the size of an amplitude. Consider {\tt P13}
in the excerpt of code above. It is composed of
\begin{verbatim}
   P13 = O11 - O12 - O13 + O14 + O20
\end{verbatim}
The {\tt O}$nn$ are in turn made up of
\begin{verbatim}
   O11 = e23*k13*k43
   O12 = e13*k23*k43
   O13 = e13*k24*k43
   O14 = e12*k31*k43
   O20 = e23*k14*k43
\end{verbatim}
and the {\tt e}$nn$ and {\tt k}$nn$ are defined as
\begin{verbatim}
   e12 = Pair[e[1], e[2]]
   e13 = Pair[e[1], e[3]]
   k14 = Pair[e[1], k[4]]
   ...
\end{verbatim}
{\tt Pair[$a$,\,$b$]} denotes the scalar product of two four-vectors, and
{\tt e[$i$]} and {\tt k[$i$]} the polarization vectors and momenta,
respectively.

To get an idea of how advantageous the introduction of abbreviations can
be, it is useful to compare the {\tt LeafCount} of the expressions in
\mma. The leaf count gives a measure of the size of an expression, more
precisely it counts the number of subexpressions or ``leaves'' on the
expression tree. {\tt P13} has a leaf count of 1 since it is just a plain
symbol. In comparison, its fully expanded contents have a leaf count of
113.

The definitions of the abbreviations can be retrieved by
{\tt Abbreviations[]} which returns a list of rules such that
\begin{verbatim}
   result //. Abbreviations[]
\end{verbatim}
gives the full, unabbreviated expression. It is of course necessary to
save the abbreviations before terminating a session, \eg with
\begin{verbatim}
   Abbreviations[] >> abbr
\end{verbatim}

\subsection{\LT}
\label{sect:looptools}

\LT\ supplies the functions needed for the numerical evaluation of the
code produced by \FC. \LT\ follows the conventions of \cite{De93}. The
complete reference to the \LT\ functions can be found in Appendix
\ref{sect:reflt}.

\LT\ consists of three parts: a Fortran library, a C++ library, and a
MathLink executable that can be used with \mma\ directly.

\subsubsection{Using \LT\ with Fortran}

To use the \LT\ functions in a Fortran program, the two files 
{\tt tools.F} and {\tt tools.h} must be included. {\tt tools.F} contains
actual Fortran code and should be included only once per program.
{\tt tools.h} contains the declarations and common blocks and must be
included in every function or subroutine in which the \LT\ functions are
used.

Before using any \LT\ function, the subroutine {\tt bcaini} must be
called. At the end of the calculation {\tt bcaexi} may be called to
obtain a summary of errors. 

A very simple program would for instance be
\begin{verbatim}
#include "tools.F"

        program simple_program
#include "tools.h"

        call bcaini
        print *, B0(1000D0, 50D0, 80D0)
        call bcaexi
        end
\end{verbatim}
Note that, as for all preprocessor commands, the {\tt\#} must stand at
the beginning of the line.

Several default values can be superseded by defining preprocessor
variables before including {\tt tools.F}, \eg the renormalization
scale can be changed with {\tt\#define MUDIM \dots} or the IR regulator
mass can be changed with {\tt\#define LAMBDA \dots}; some more technical
options are described in the manual \cite{FCLTGuide}.

\subsubsection{Using \LT\ with C++}

To use the \LT\ functions in a C++ program, the file {\tt ctools.h} must
be included. Similar to the Fortran case, before making the first call to
any \LT\ function, {\tt bcaini()} must be called and at the end
{\tt bcaexi()} may be called to get a summary of errors.

In C++, the same simple program looks like
{\samepage
\begin{verbatim}
#include <fstream.h>
#include "ctools.h"

main()
{
  bcaini();
  cout << B0(1000., 50., 80.) << endl;
  bcaexi();
}
\end{verbatim}}

The renormalization scale and the IR regulator mass can be changed with
the functions {\tt set\symbol{95}mudim} and {\tt set\symbol{95}lambda},
respectively.

\subsubsection{Using \LT\ with \mma}

The \mma\ interface is probably the simplest to use:
\begin{verbatim}
In[1]:= Install["bca"]

Out[1]= LinkObject[bca, 1, 1]

In[2]:= B0[1000, 50, 80]

Out[2]= -4.40593 + 2.70414 I
\end{verbatim}
One-loop functions containing non-numeric arguments (\eg
{\tt B0[1000, MW2, MW2]}) remain unevaluated.

Again, the renormalization scale and the IR regulator mass may be changed
with the functions {\tt Mudim} and {\tt Lambda}, respectively.

\section{An application to the Standard Model}
\label{sect:example}

We demonstrate in this section the basic steps of a real one-loop
calculation using our packages. For an example, we choose the process
\ZZZZ\ in the electroweak Standard Model \cite{DeDH97}.

A one-loop calculation generally includes
\begin{center}
\begin{tabular}{l|l|l}
Step & Program & typical CPU time \\ \hline
1. Generate Diagrams & \FA & 3 min. \\
2. Simplify analytically & \FC & 10 min. \\
3. Produce Fortran code & ({\sl NumPrep}) & 3 min. \\
4. Compile with driver program & ({\tt num.F}) & 7 min.
\end{tabular}
\end{center}
The quoted execution times are for the full one-loop \ZZZZ\ calculation
(about 500 diagrams) as run on a standard Pentium PC under Linux.

The numerical evaluation of the \FC\ output is not fully automated.
The reason is that at one-loop level already many features appear which
must be treated differently from process to process. The two programs
{\sl NumPrep} and {\tt num.F} are supplied with the demo code to show how
the numerical evaluation can be done. Although both were designed for
$2\to 2$ gauge-boson scattering processes, it should not be too difficult
to adapt them at least to other $2\to 2$ processes. (Basically, one has to
supply the kinematics in {\tt num.F}.)

In the following we present only a brief but characteristic excerpt of
code for each step. The complete demo code is contained in the online
distribution of \FC.

\begin{description}
\item[Step 1:] Here the generation of the self-energy diagrams is
demonstrated:
\begin{verbatim}
<< FeynArts.m

tops = CreateTopologies[ 1, 2 -> 2,              (* 1 loop, 2 -> 2 *)
  ExcludeTopologies ->                (* create only self-energies *)
    {Tadpoles, WFCorrections, Triangles, AllBoxes} ]

inss = InsertFields[ tops,
  {V[2], V[2]} -> {V[2], V[2]},                        (* V[2] = Z *)
  Model -> "SM", InsertionLevel -> {Particles},
  Restrictions -> {NoGeneration2, NoGeneration3} ]

amps = CreateFeynAmp[inss];
ToFA1Conventions[amps] >> zzzz.self.amp
\end{verbatim}
This code excerpt shows a typical application of \FA\/: create the
topologies, insert fields into them ({\tt V[2]} is a Z boson in \FA\
lingo), and create the amplitudes, \ie apply the Feynman rules. The
function {\tt ToFA1Conventions} converts the full \FA\ 2.2 format back
into the simpler \FA\ 1 format which is needed by \FC.

\item[Step 2:] The so-created self-energy diagrams are then simplified
with \FC. Calculating in $D$ and 4 dimensions yields equivalent results in
the case of $\rm ZZ\to ZZ$, hence we have omitted the explicit definition
of {\tt \$Dimension} here (the default is {\tt D}).
\begin{verbatim}
<< FormCalc.m

ProcessFile["zzzz.self.amp", "self"];
Abbreviations[] >> abbr
\end{verbatim}

\item[Steps 3 and 4:] The \FC\ results need to be converted into a
Fortran program. (The numerical evaluation could, in principle, be done in
\mma\ directly, but this becomes very slow for large amplitudes.) The
simplest way to do this in \mma\ is something like
{\tt FortranForm[result] >> file.f}. With our demo code we supply
a much more sophisticated program called {\sl NumPrep} which goes well
beyond simple translation into Fortran code. For example, it groups the
one-loop integrals into angle-dependent and -independent integrals, so
that the latter need to be calculated only once \eg when integrating over
the angle at a fixed energy.

{\sl NumPrep} is too complex to describe here in detail. Instead, we give
an example of how the final Fortran code looks like:
\begin{verbatim}
#include <defs.h>
        double complex function self()
        implicit logical (a-s,u-z)
        implicit double complex (t)
#include <vars.h>
        self = reso**2*(-5.7477663296703327914881096651780*ab61*O3 -
     -     3.696762828539980268636174052200372*ab62*O3 +
     -     ab59*(14.787051314159921074544696208801489 +
     -        0.00022228809903668380734374191468819973*MH2)*O3 +
     -     ab58*(22.991065318681331165952438660711951 +
     -        0.0004445761980733676146874838293763995*MH2)*O3 +
     -     0.0006668642971100514220312257440645992*ab60*MH2*O3 +
     -     0.0020005928913301542660936772321937976*ab15*MH2**2*
     -      O3 - ab14*O3*(-215176.58454926418740557299134203950 -
     -        0.00022228809903668380734374191468819973*MH2**2 +
     -        7.393525657079960537272348104400745*S) - ...
\end{verbatim}
Admittedly, this code looks ugly to the human eye, but note that it is
highly optimized for numerical evaluation: apart from the {\tt O3} which
is one of the abbreviations already introduced by \FC, the one-loop
integrals have been replaced by variables (\eg {\tt ab}{\it nn} for A
and B functions). Also, the possibly resonant Higgs propagator
$1/(s - \MH^2)$ has been replaced by the variable {\tt reso} so that it
can be treated more easily in Fortran. These are just examples of what one
can do with the \FC\ output.

Of course, {\sl NumPrep} produces also the code to calculate the
abbreviations and a {\tt Makefile} to conveniently compile the code.
The Fortran code produced by {\sl NumPrep} needs in addition a driver
program which supplies it with the necessary parameters, kinematics, etc.
This driver program is called {\tt num.F} and is included in the demo.
\end{description}

\section{Computer requirements and availability}
\label{sect:availability}

\FC\ and \LT\ should compile and run without change on any Unix-based
platform. In particular, they have been tested on DEC Alpha, HP 9000,
and Linux. \FC\ needs \mma\ 2.2 or above including the MathLink compiler
({\tt mcc}) and \FO\ 2 or above.\footnote{%
\FC\ runs also with \FO\ 1, but cannot fully simplify spinor chains with
external fermions then.}
\LT\ requires a Fortran-77 compiler, the GNU make utility, and the GNU C
and C++ compilers ({\tt gcc}, {\tt g++}).

The programs \FC\ and \LT\ can be obtained via WWW from \\
\hspace*{2em}{\tt http://www-itp.physik.uni-karlsruhe.de/formcalc}
and \\
\hspace*{2em}{\tt http://www-itp.physik.uni-karlsruhe.de/looptools},
respectively.\\
The packages contain a comprehensive manual giving installation
instructions and a detailed description of every function.

\FA\ is available from \\
\hspace*{2em}{\tt ftp://ftp.physik.uni-wuerzburg.de/pub/hep/index.html}.

\section{Acknowledgements}

We thank Ansgar Denner for important discussions. T.H.\ thanks the
University of Granada for the hospitality during preparation of this
work. This work has been partially supported by DFG, under contract number
Ku 502/8--1, and by CICYT, under contract number AEN96-1672, Junta de
Andalucia, FQM101 and Ministerio de Educacion y Cultura.

\begin{appendix}

\section{Reference of the \LT\ functions}
\label{sect:reflt}

\subsection{One-point functions}

\begin{center}
\begin{tabular}{|l|l|} \hline
Function call & Description \\ \hline
{\tt A0(ms)} &
	one-point function \\
\hline
\end{tabular}
\end{center}

The real argument {\tt ms} ({\tt double precision} in Fortran) is the
mass squared:
\begin{equation}
\begin{array}{lll}
\begin{array}{l}
\begin{aligned}
\mathtt{ms} &= m^2
\end{aligned}
\end{array}
 & \quad &
\begin{array}{l}
\begin{picture}(100,40)(10,23)
\Line(20,40)(50,40)
\Vertex(50,40){2}
\CArc(70,40)(20,0,360)
\Text(96,40)[cl]{$m$}
\end{picture}
\end{array}
\end{array}
\end{equation}

\subsection{Two-point functions}

\begin{center}
\begin{tabular}{|l|l|} \hline
Function call & Description \\ \hline
{\tt B0(ps, m1s, m2s)} &
	scalar two-point function \\
{\tt B1(ps, m1s, m2s)} &
	coefficient of $p_\mu$ \\
{\tt B00(ps, m1s, m2s)} &
	coefficient of $g_{\mu\nu}$ \\
{\tt B11(ps, m1s, m2s)} &
	coefficient of $p_\mu p_\nu$ \\
\hline
\end{tabular}
\end{center}

All arguments are real ({\tt double precision} in Fortran) and are
related to the momenta and masses as follows:
\begin{equation}
\begin{array}{lll}
\begin{array}{l}
\begin{aligned}
\mathtt{ps} &= p^2 \\[1ex]
\mathtt{m1s} &= m_1^2 \\[1ex]
\mathtt{m2s} &= m_2^2
\end{aligned}
\end{array}
 & \quad &
\begin{array}{l}
\begin{picture}(140,60)(0,13)
\ArrowLine(20,40)(50,40)
\ArrowLine(90,40)(120,40)
\CArc(70,40)(20,0,360)
\Vertex(50,40){2}
\Vertex(90,40){2}
\Text(16,40)[cr]{$p$}
\Text(125,40)[cl]{$p$}
\Text(72,63)[bc]{$m_1$}
\Text(72,15)[tc]{$m_2$}
\end{picture}
\end{array}
\end{array}
\end{equation}

\subsection{Derivatives of two-point functions}

\begin{center}
\begin{tabular}{|l|l|} \hline
Function call & Description \\ \hline
{\tt DB0(ps, m1s, m2s)} &
	derivative of {\tt B0} \\
{\tt DB1(ps, m1s, m2s)} &
	derivative of {\tt B1} \\
{\tt DB00(ps, m1s, m2s)} &
	derivative of {\tt B00} \\
{\tt DB11(ps, m1s, m2s)} &
	derivative of {\tt B11} \\
\hline
\end{tabular}
\end{center}

All derivatives are with respect to the momentum squared. The arguments
are as in the case of the two-point functions.

\subsection{Three-point functions}

\begin{center}
\begin{tabular}{|l|l|} \hline
Function call & Description \\ \hline
{\tt C0(p1s, p2s, p1p2s, m1s, m2s, m3s)} &
	scalar three-point function \\
{\tt C0i(id, p1s, p2s, p1p2s, m1s, m2s, m3s)} &
	three-point tensor coefficients \\
\hline
\end{tabular}
\end{center}

Except for the {\tt id} all arguments are real ({\tt double precision}
in Fortran) and are related to the momenta and masses as follows:
\begin{equation}
\begin{array}{lll}
\begin{array}{l}
\begin{aligned}
\mathtt{p1s} &= p_1^2 \\[1ex]
\mathtt{p2s} &= p_2^2 \\[1ex]
\mathtt{p1p2s} &= (p_1 + p_2)^2 \\[1ex]
\mathtt{m1s} &= m_1^2 \\[1ex]
\mathtt{m2s} &= m_2^2 \\[1ex]
\mathtt{m3s} &= m_3^2
\end{aligned}
\end{array}
 & \quad &
\begin{array}{l}
\begin{picture}(155,140)(0,13)
\ArrowLine(20,20)(40,40)
\ArrowLine(20,140)(40,120)
\ArrowLine(136,80)(105,80)
\Line(40,40)(40,120)
\Line(105,80)(40,40)
\Line(40,120)(105,80)
\Vertex(40,40){2}
\Vertex(105,80){2}
\Vertex(40,120){2}
\Text(16,140)[br]{$p_1$}
\Text(141,80)[cl]{$p_2$}
\Text(16,20)[tr]{$p_3$}
\Text(36,80)[cr]{$m_1$}
\Text(75,101)[bl]{$m_2$}
\Text(75,58)[tl]{$m_3$}
\end{picture}
\end{array}
\end{array}
\end{equation}

{\tt C0i} is a generic function for all three-point tensor coefficients.
A specific coefficient is selected with the {\tt id} argument:
\begin{equation}
\begin{aligned}
\text{\tt C0i(cc0,\,\dots)} &= C_0(\ldots) \\
\text{\tt C0i(cc00,\,\dots)} &= C_{00}(\ldots) \\
\text{\tt C0i(cc112,\,\dots)} &= C_{112}(\ldots) \quad \text{etc.}
\end{aligned}
\end{equation}
Since the indices are symmetric, the identifiers are assumed to be
ordered, \ie there is only {\tt cc122} but not {\tt cc212}.

\subsection{Four-point functions}

\begin{center}
\begin{tabular}{|l|l|} \hline
Function call & Description \\ \hline
{\tt D0(p1s, p2s, p3s, p4s, p1p2s, p2p3s,} &
  scalar four-point function \\
{\tt ~~~m1s, m2s, m3s, m4s)} & \\
{\tt D0i(id, p1s, p2s, p3s, p4s, p1p2s, p2p3s,} &
  four-point tensor coefficients \\
{\tt ~~~m1s, m2s, m3s, m4s)} & \\
\hline
\end{tabular}
\end{center}

Except for the {\tt id} all arguments are real ({\tt double precision}
in Fortran) and are related to the momenta and masses as follows:
\begin{equation}
\begin{array}{lll}
\begin{array}{l}
\begin{aligned}
\mathtt{p1s} &= p_1^2 \\[1ex]
\mathtt{p2s} &= p_2^2 \\[1ex]
\mathtt{p3s} &= p_3^2 \\[1ex]
\mathtt{p4s} &= p_4^2 \\[1ex]
\mathtt{p1p2s} &= (p_1 + p_2)^2 \\[1ex]
\mathtt{p2p3s} &= (p_2 + p_3)^2 \\[1ex]
\mathtt{m1s} &= m_1^2 \\[1ex]
\mathtt{m2s} &= m_2^2 \\[1ex]
\mathtt{m3s} &= m_3^2 \\[1ex]
\mathtt{m3s} &= m_4^2
\end{aligned}
\end{array}
 & \quad &
\begin{array}{l}
\begin{picture}(130,125)(0,0)
\ArrowLine(5,10)(30,30) 
\ArrowLine(5,115)(30,95)
\ArrowLine(120,115)(95,95)
\ArrowLine(120,10)(95,30)
\Line(95,30)(30,30)
\Line(30,30)(30,95)
\Line(30,95)(95,95)   
\Line(95,95)(95,30)
\Vertex(30,30){2}
\Vertex(30,95){2}
\Vertex(95,30){2}
\Vertex(95,95){2}
\Text(0,115)[r]{$p_1$}
\Text(125,115)[l]{$p_2$}
\Text(125,7)[l]{$p_3$}
\Text(0,7)[r]{$p_4$}
\Text(25,62)[r]{$m_1$}
\Text(62,100)[b]{$m_2$}
\Text(100,62)[l]{$m_3$}
\Text(62,24)[t]{$m_4$}
\end{picture}
\end{array}
\end{array}
\end{equation}

{\tt D0i} is a generic function for all four-point tensor coefficients. A
specific coefficient is selected with the {\tt id} argument:
\begin{equation}
\begin{aligned}
\text{\tt D0i(dd0,\,\dots)} &= D_0(\ldots) \\
\text{\tt D0i(dd00,\,\dots)} &= D_{00}(\ldots) \\
\text{\tt D0i(dd1223,\,\dots)} &= D_{1223}(\ldots) \quad \text{etc.}
\end{aligned}
\end{equation}
Again, since the indices are symmetric, the identifiers are assumed to be
ordered, \ie there is only {\tt dd1223} but not {\tt dd3212}.

\end{appendix}

\begin{flushleft}

\end{flushleft}

\end{document}